# Development of a pulse oximeter robust to measurement errors, with the ability to estimate heart rate and transmit data to smartphones

Hosna Ghandeharioun
Assistant professor
Department of Electrical & Biomedical Engineering
Khorasan Institute of Higher Education
Mashhad, Iran
https://orcid.org/0000-0003-1177-3394

Ali Akbari
B.S. Graduate
Department of Electrical & Biomedical Engineering
Khorasan Institute of Higher Education
Mashhad, Iran
Email: :mraliak1379@gmail.com

*Abstract*—Accurate and real-time monitoring of saturated oxygen level of the blood is an important clinical issue gaining great attention in recent years and during COVID 19 pandemic. Monitoring the patients` ventilation and respiration dynamic is widespread and has been adopted as a standard for anesthesia, neonatal care, and post-operative recovery as well. In this paper the fundamentals of current pulse oximeter devices are reviewed and development of a prototype is explained. Our system has two red and infrared light sources radiating to the tissue. The amount of absorbed and transmitted energy is measured by the photodetector, and finally the amount of oxygen content of the blood is estimated based on these values. We used MAX30100 module. Our system has two advantages relating to similar devices; robustness to errors due to power supply variations, ambient light and motion artifacts. Our system has also the ability to estimate heart rate and transmit the data to a smart phone. This makes our proposed system a potentially good hardware for home-monitoring of blood oxygen and respiration efficiency if collaborated with a good mobile application for detecting blood de-saturations.

***Keywords-non-invasive oximetery, pulse oximetry, home-based health monitoring, mobile health (mHealth), Ardoino, MAX30100***

## I. Introduction

Fast and accurate measurement of saturated blood oxygen and other biological signals plays a crucial role in clinical applications. Nowadays for noninvasive measurement of saturated oxygen level of the blood pulse oximetry is used. The optical principle of pulse oximetry technique is the Beer-Lambert law [1]. This law states that the absorption of light of a given wavelength passing through a non-absorbing solvent containing an absorbing solute, is proportional to the product of the solute concentration, the light path length, and an extinction coefficient. The Beer-Lambert law can readily be applied in a laboratory setting. However, it must be modified for in vivo situations to overcome the obstacles associated with interference from tissue and pulsatile flow. This modification involves measuring absorbance at two different wavelengths, one to detect oxyhemoglobin and the other to detect deoxyhemoglobin finally asses the level of peripheral arterial oxygen saturation ($SpO_2$) [2].

Patients with sleep apnea, chronic obstructive pulmonary disease (COPD), asthma, pneumonia, anemia, lung cancer, congestive heart failure, or any other cardio-respiratory problem use portable pulse oximeters to continuously monitor their $SpO_2$. In pulmonary diseases, pulse oximeters are also used in certain time intervals after the intake of a new medication to evaluate the efficiency the treatment. During oxygen therapy and even athletes in hard exercises use this device to monitor their ventilation.

The $SpO_2$ levels around 95% and more are considered normal, while values less than 92% are certainly called hypoxia (shortage of supplied oxygen in tissue). Long-term hypoxia has serious effects on body cells. Hence the monitoring of arterial oxygen saturation and abrupt treatment of hypoxia is of great importance [7].

*A. Research History*

In 1931, Nicolai investigated the quantitative spectrophotometry of light transmitted through human tissues. In 1934, Kramer precisely measured the oxygen saturation of blood flowing through cuvettes. Measurement of the spectrum of undiluted hemolyzed and non-hemolyzed blood was accomplished by Drabkin and Austin in 1935. In 1940 Squires passed red and infrared light through the finger web for the continuous monitoring of oxygenation. At the same time Millikan coined the term oximeter and developed the Millikan oximeter. Seven years later Wood developed an ear oximeter accompanied with a pure oxygen capsule to retain complete oxygen saturation. In 1960 the first bench "CO-oximeter" was developed with the ability to distinguish between hemoglobin, carboxyhemoglobin and methemoglobin. In 1964, Shaw developed the eight-wavelength ear oximeter. This device was introduced to market by Hewlett-Packard later. In 1971, Aoyagi used the pulsatility of the absorption signal to separate absorption due to the arteries from the other tissues and three years later he developed the prototype pulse oximeter using incandescent light source, filters, and analog electronics. The advent of the first commercially available pulse oximeter was in 1975. Diab and Kiani from Masimo corporation in 1989, developed a new pulse oximeter utilizing parallel processing engines and adaptive filters to the measurement signal .It enabled direct calculation of arterial oxygen saturation and pulse rate and is not bounded by a conventional "red over infrared" ratio approach, hence, eliminated the problems of motion artifact, low peripheral perfusion and most low signal-to-noise situations. The commercial version named Masimo SET was introduced to market in 1996 [5]. The only Iranian commercial pulse oximeter for clinical use is made by Mapva, however several other companies has made efforts [10].

### B. Problem Definition

The main problem in all the non-commercial prototypes is fluctuation of the measurements due to power supply and photodiode light emitting power alterations. Motion artifact and ambient noise are other sources of measurement errors [2], [3], [5], [7].

In this paper the pulse oximeter is developed with MAX30100 module. Our device has both the ability of estimating heart rate and transmitting the measurement data to a smart phone. Additionally, by carefully regulating the power supply, measurement fluctuations due to power supply is deleted.

## II. MATERIALS AND METHODS

### A. Principles of pulse oximetry

As stated earlier, the most common non-invasive technique for monitoring oxygen saturation of the blood is pulse oximetry. It exploits spectrophotometry to determine the proportion of oxyhemoglobin. Red cells have different hemoglobin compounds. 99% of blood hemoglobin is de-oxygenated hemoglobin (Hb) and oxyhemoglobin ($HbO_2$). The biggest portion of blood is made of water, Hb and $HbO_2$. In wavelengths less than 1000nm, the extinction coefficient [1] of water is negligible, hence in these wavelength incident light can give us information on the ratio of Hb and $HbO_2$.

A common pulse oximeter consists of two light-emitting diodes and two sensors. One diode emits nearly red light (of 660 nm wave length). This wavelength is less absorbed and more reflected by the fully oxygenated hemoglobin (i.e. saturated hemoglobin/ oxyhemoglobin) whereas the other diode emits infrared light (wave length of 940 nm) which is less absorbed and more reflected by the fully deoxygenated adult hemoglobin (i.e. de-saturated hemoglobin). Refer to figure 1 for a detailed information. The ratio of light absorbance between oxyhemoglobin and the sum of oxyhemoglobin plus deoxyhemoglobin is detected and compared with previously calibrated direct measurements of arterial oxygen saturation ($SaO_2$) to establish an estimated measure of peripheral arterial oxygen saturation ($SpO_2$) [2].

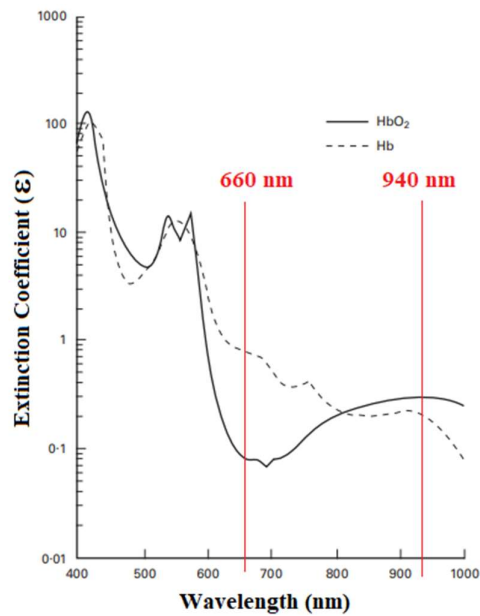

Figure 1. Absorption spectra of oxygenated (HbO2) and deoxygenated (Hb) forms of adult hemoglobin [1]

The parameters that are measured in pulse oximetry for quantitative assessment of oxygen level of the blood are partial pressure of oxygen ($PO_2$), the percentage of oxygen saturation ($PO_2$) n and peripheral arterial oxygen saturation ($SpO_2$) [6].

### B. Hardware implementation and tuning the power supply circuit

The MAX30100 module has I2C serial output. Data is transmitted through messages and each message is divided into several data frames. A message sample in I2C standard in shown in figure 2 [8].

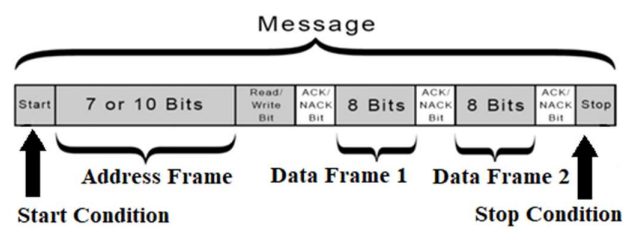

Figure 2. Message configuration in I2C standard [9]

The MAX30100 module has power supply range of 3.3 -5V. This module has an exclusive installation library that makes it easy-to-use. A low-power pulse oximeter and heart rate sensor is embedded in the module.

The supply power input of this module can be 3.3-5V due to its 1.8V regulator. Three pull-up resistances of 4.7 kΩ values are mounted on the board on SCL, SDA and INT pins.

To set up this module with a microcontroller or Arduino, the logic high level of 5V should be provided, otherwise the serial I2C port is not connected and the MAX30100 light is not lit up. To reach this goal, the pull up resistors should be carefully separated from the board. However, after connecting the module to microcontroller or Arduino, 2 pull-up resistances are to be connected between Vcc and SDA/SCL pins. Since pull-ups are exposed to a voltage more than 1.8V, the module power supply should not exceed 4.3V. See figure 3.

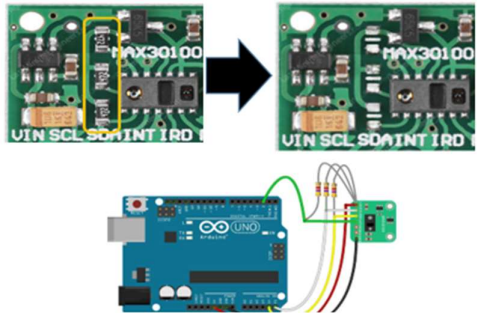

Figure 3. Tunning the power supply circuit for MAX30100

It is also possible to disconnect the last pull-up resistance from the 1.8V regulator and connect it directly to the module without unmounting the resistors from the board. However, this approach does not work here since MAX30100 turns on with logic high voltage level (i.e. 5V).

To regulate the module`s power supply the step-down DC-DC converter LM2596 is used. This regulator can supply a load up to 3 A. to compensate for the wire`s loss we applied 5V to the regulator LM2596 from Arduino to gain 4.2V after all losses.

The MAX30100 board is placed in a frame and the whole system is put in a case after primary testing to reduce the noise due to ambient light. Figures 4-a to 4-c show the circuit schematics and its hardware realization.

### C. Software development and data transmission to smart phones

Uploading the codes [11] using Arduino software version 1.8.15 and setting the bud rate to 1152, the measured $SpO_2$ and heartrate are exhibited on the LCD and the serial monitor of Arduino software. In this project the UART cable is employed to upload codes on the Arduino. Data transmission to the smart phone is also accomplished via this cable using a USB-type C convertor. To see the measurements on the phone, Serial USB terminal should be installed on the phone and the bud rate set to the right value. The eventual view of the system is shown in figure 5.

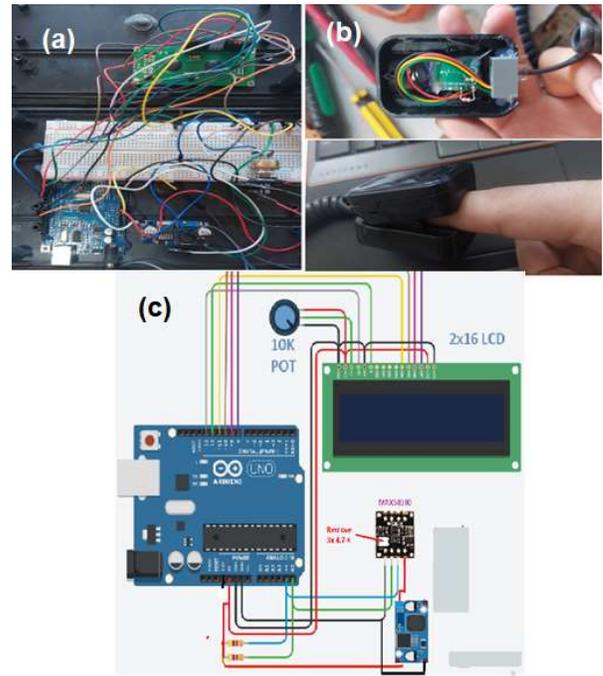

Figure 4. (a) System hardware, (b) MAX30100 placed in a frame, (c) Circuit schematics: Vin of MAX30100 connected to 4.2V output pin of LM2596, GND of MAX30100 connected to GND of LM2596, SCL of MAX30100 connected to a5 of Arduino, SDA of MAX30100 connected to a4 of Arduino

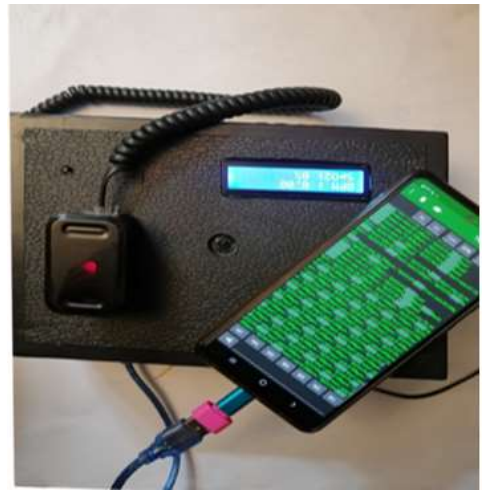

Figure 5. The eventual view of the packed system

### III. RESULTS

The measured values of $SpO_2$ by our designed system are shown in figure 6. Simultaneously an ONYX NONIN [II] 9560 senor is also measuring the $SpO_2$ with sampling frequency of 1 Hz. This sensor is used as a bench mark to evaluate measurements of our designed system. As it is demonstrated in figure 6, the values of our system and those of ONYX NONIN [II] 9560 are concordant, and according to table 1, the average measurement errors for both subjects are less than 0.1% compared to the bench mark pulse oximeter It is worth noting that to calculate the error for each measurement, the ONYX NONIN $^{II}$ 9560 measured value is considered as the true value ($x_{i,true}$) and our system value is considered as the measured value ($x_{i,measured}$), finally the average of errors is reported for twenty consecutive measurements;;

$$\text{Average Measurement Error} = \sum_{i=1}^{20} \frac{x_{i,true} - x_{i,measured}}{x_{i,true}} \quad (1)$$

### IV. CONCLUSION

Considering the fact that heart rate and SpO2 are two important biological signals, their real-time measurement is very significant. The proposed system is a sample device with good measurement accuracy and robustness against power supply fluctuations and ambient light. It has the ability to transmit data to smart phone for afterwards processing of a possible health-monitoring application.

Instead of using UNO Arduino, a microprocessor of the same family or an ESP8266 module can be employed to reduce size considerably. To improve better presentation, an OLED monitor can be used and to transmit data to a nearby smartphone instead of a UART cable, a WIFI module can be exploited. However, using WIFI modules is at the expense of higher power consumption. It is worth-noting that providing the measurement data for a smart phone in real-time is a valuable potential of the system. The system can easily collaborate with an application of SpO2 monitoring for home-based usage.

TABLE I.   AVERAGE MEASUREMENT ERROR IN 20 SESSIONS

| Subject/Gender | Average Measurement Error |
|---|---|
| A/Male | *0.07%* |
| H/Female | 0.03% |

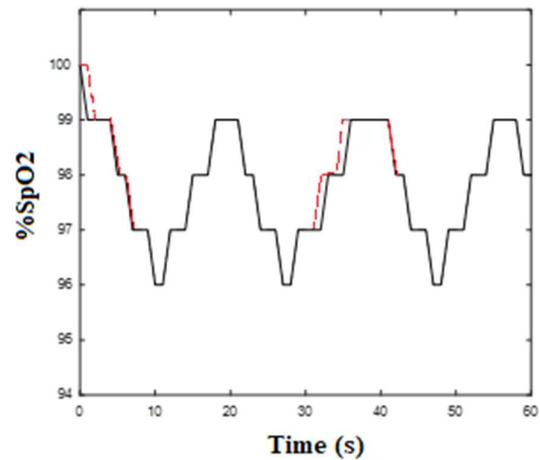

Figure 6.   The measured SpO2 values in one minute by our proposed sytem (solid line,black) versus the ONYX NONIN $^{II}$ 9560 (dotted line, red).Mostof the two graghs have overlap.

ACKNOWLEDGMENT

The authors wish to thank Khorasan Institute of Higher Education for its sponsorship.